\documentstyle[preprint,eqsecnum,aps]{revtex}
\def\lsim{\:\raisebox{-0.5ex}{$\stackrel{\textstyle<}{\sim}$}\:}
\def\gsim{\:\raisebox{-0.5ex}{$\stackrel{\textstyle>}{\sim}$}\:}
\begin{document} \draft
\preprint{TIFR/TH/95-28}

\title{Parton Gas Model for the Nucleon Structure Functions}

\author{R.S. Bhalerao\cite{byline}}
\address{Theoretical Physics Group\\
Tata Institute of Fundamental Research\\
Homi Bhabha Road, Colaba, Bombay 400 005, India}

%\date{\today}
\maketitle

\begin{abstract}
A phenomenological model for the nucleon structure functions is
presented. Visualising the nucleon as a cavity filled with parton gas
in thermal equilibrium and parametrizing the effects due to the
finiteness of the nucleon volume, we obtain a good fit to the data on
the unpolarized nucleon structure functions.
\end{abstract}


\narrowtext

Recent experiments have revealed some remarkable features of the
nucleon structure functions $F_2^{p,n}$. Data on deep inelastic
scattering of muons off proton and deuteron targets \cite{nmcgsr}
show that the quark sea in the nucleon is not flavor-symmetric,
$\bar u(x) \neq \bar d(x)$; the Gottfried sum \cite{gsr} $S_G \equiv
\int \left( F^p_2 - F^n_2 \right) (dx/x)$, at $Q^2 = 4$ GeV$^2$,
has the value $0.235 \pm 0.026$ compared to the usual quark model
prediction of 1/3. This result has been confirmed by the observed
asymmetry in Drell-Yan production of dileptons in $pp$ and $pn$
collisions \cite{NA51}. Most
notably, the HERA electron-proton scattering data \cite{hera} reveal a
rapid rise of the proton structure function $F^p_2 (x)$ as $x$
decreases. Indeed over a wide range of small $x$, data from the
various groups \cite{hera,nmcbcdms}, for fixed $Q^2$,
are all well described by a single
inverse power of $x$. Figure 1 is a log-log plot of the data on
$F_2^p(x)/x$ (the combination that enters $S_G$) versus $x$. We see
that, for fixed $Q^2$, the data fall on straight lines defined by
\begin{equation}
{F^p_2(x) \over x}={c\over x^m}~, ~(0.0004 \lsim x \lsim 0.2).
\eqnum{1}
\end{equation}
For instance, at $Q^2=15$ GeV$^2$, the best-fit parameters are
$c=0.229\pm 0.005$ and $m= 1.22 \pm 0.01$ \cite{fn}.

Global fits to the nucleon structure data involve parametrizing the
various parton densities at some low $Q^2$ and evolving them to higher
values of $Q^2$ relevant to observations. The fits so obtained
\cite{params} have very high precision but contain several (typically
$\sim $15-20) arbitrary parameters and provide little physical
insight into the structure of the nucleon. On the other hand,
phenomenological models could give us some valuable clues into the
physics of parton distributions in the nucleon. From this point of
view the statistical models of the nucleon structure functions
\cite{gasms} have been quite interesting due to their intuitive appeal and
simplicity.

We present here a phenomenological model for the unpolarized nucleon
structure functions by regarding the nucleonic contents as
constituting a gas of noninteracting partons in thermal
equilibrium. An attractive feature of this general framework is the
natural explanation of the violation of the Gottfried sum rule: the
excess of $u$-quarks over $d$-quarks in the proton implies unequal
chemical potentials, hence unequal $u\bar u$ and $d\bar d$ seas, which
leads to $S_G\neq 1/3$. However, the ensuing structure function
$F_2(x)$ vanishes like $x^2$ as $x \rightarrow 0$, in violent conflict
with the data. To remedy this we invoke corrections arising from the
finiteness of the nucleon volume, by multiplying the parton density of
states by a factor having inverse powers of the radial dimension,
$[1+{\cal O}(1/ R^\delta )]$. We find that the small-$x$ rise and
other features of the nucleon structure functions are reproduced quite
well.

\medskip

\noindent {\it The Model}

We picture the nucleon (mass $M$) to consist of a gas of massless
partons (quarks, antiquarks and gluons) in thermal equilibrium at
temperature $T$ in a spherical volume $V$ with radius $R$. We consider
two frames, the proton rest frame and the infinite-momentum frame
(IMF) moving with velocity $-v (\simeq -1)$ along the common $z$
axis. Our interest lies in the limit when the Lorentz factor $\gamma
\equiv (1-v^2)^{-1/2} \rightarrow \infty$. The invariant parton number
density in phase space \cite{degroot} is given by (quantities in the
IMF are denoted by the index $i$)
\begin{eqnarray}
{dn^i \over d^3p^i~d^3r^i}={dn\over d^3p~d^3r}
&=& {g\over (2\pi )^3} \left[{1\over {\exp[\beta (E-\mu)]\pm 1}}\right]
\nonumber \\
&\equiv & f(E), \eqnum{2}
\end{eqnarray}
where $\beta \equiv T^{-1}$,
$g$ is the degeneracy ($g=16$ for gluons and $g=6$ for $q$ or
$\bar q$ of a given flavor), $(E,\bf p)$ is the parton four-momentum
in the proton rest frame and $f(E)$ is the usual distribution for
noninteracting fermions or bosons. In terms of the Bjorken scaling
variable $x =p^i_z/(Mv \gamma )$, the phase space element can be
expressed as
\begin{eqnarray}
d^3p^i~d^3r^i
&&=2\pi p^i_Tdp^i_T(Mv \gamma dx){d^3r\over \gamma}\nonumber \\
&&=2\pi [Mxv^3+{Ev\over \gamma ^2}]dE~Mdx~d^3r,\nonumber
\end{eqnarray}
where for $\gamma \rightarrow \infty $ the expression in square
brackets becomes $Mx$. For fixed $x$ the parton energy $E$ varies
between the kinematic limits $Mx/2< E < M/2$, where the lower limit is
attained when $p^i_T = 0$. Consequently the parton number distribution
$dn^i/dx$ in the IMF is simply proportional to an
integral of the {\it rest-frame} distribution $f(E)$:
\begin{equation}
dn^i/dx =2\pi V M^2 x \int ^{M/2}_{xM/2} dE ~~f(E),\eqnum{3}
\end{equation}
where the factor $V$ results from $d^3r$ integration. The structure
function $F_2(x)$ is given by
\[
F_2(x) = x \sum_q e_q^2 \left[{dn_q^i \over dx} + {dn_{\bar q}^i \over dx}
\right].
\]
The number distribution vanishes linearly as $x \rightarrow 0$ (and also as
$x \rightarrow 1$) and leads to the behavior of the structure function
$F_2(x) \sim x^2$ at small $x$, which disagrees with the observations
noted in Eq. (1).

In order to obtain the rise of $F_2^p(x)$ at small $x$, we shall modify
the model to reflect effects arising from the finiteness of the
nucleon volume $V$. Various studies of finite-size corrections (FSC)
show that they are sensitive to the precise shape and size of the
enclosure, the type of boundary conditions imposed on the wave
function, and to the details such as whether the particles are
strictly massless, whether chemical potentials are nonzero, etc.
\cite{fse}. Moreover, these studies invariably involve some simplifying
assumptions and thus their use is difficult to justify in the present
context.

In keeping with the phenomenological nature of the model, we have
chosen to parametrize the correction due to the finiteness of the
nucleon volume. This is implemented through the use of the dimensionless
combination $1/(ER)$. We have chosen two alternative forms
of parametrization, a form
prompted by the empirical observation in Eq. (1):
\begin{equation}
\Phi_1 = 1+{B\over (ER)^\delta}~, \eqnum{4}
\end{equation}
and a general power series expansion in the variable $1/(ER)$:
\begin{equation}
\Phi_2 = 1+{a\over ER}+{b\over (ER)^{2}}+ ...~,
\eqnum{5}
\end{equation}
where $B,~\delta(>0),~a,~b,~...$ are arbitrary constants. We multiply the
integrand in Eq. (3) by the function $\Phi~(=\Phi_1 ~\rm{or}~ \Phi_2)$
in order to incorporate the finite-volume
effects in our model.

The model described above is assumed to hold at a certain input momentum
scale $Q_0^2$, and if necessary can be evolved to higher $Q^2$ by means
of the standard techniques in quantum chromodynamics (QCD).
To complete the statement of the model, we demand the thermal parton
distributions to obey the following {\it three} constraints
at the input scale. The
constraints on the net quark numbers in the proton are $n_u - n_{\bar u} =
2$ and $n_d - n_{\bar d} = 1$, i.e.,
\widetext
\begin{eqnarray}
{V M^2 \over (2\pi)^2} \int ^1_0 dx~ x~\int^{M/2}_{xM/2}dE&&
\left\{{6 \over \exp \big[\beta (E - \mu _\alpha)\big] +1} -
{6 \over \exp\big[\beta (E + \mu _\alpha)\big] +1} \right\}
\Phi(ER)\nonumber \\
&=& n_\alpha -n_{\bar\alpha}.~~~~~~~~~~(\alpha = u,d) \eqnum{6}
\end{eqnarray}
Obviously, chemical potentials for heavy flavors are necessarily zero.
As regards the third constraint, we assume that the
longitudinal momentum fractions in the $u,~d$ flavors and the gluons
add up to unity:
\begin{eqnarray}
{V M^2 \over (2\pi)^2} \int ^1_0
dx~x^2 \int^{M/2}_{xM/2} dE~~ \bigg\{ {6 \over \exp \big[\beta (E - \mu
_u)\big] +1} &+& {6 \over \exp\big[\beta (E+\mu _u)\big] +1} + \nonumber \\
{6 \over \exp \big[\beta (E-\mu _d)\big] +1} + {6 \over
\exp\big[\beta (E + \mu _d)\big] +1} &+&
{16 \over \exp(\beta E) -1} \bigg\} {\Phi }(ER)~ = 1~. \eqnum{7}
\end{eqnarray}
The quark flavors $s,~c,~...$ which are not introduced in Eq. (7)
show up at higher $Q^2$ as a result of QCD evolution.
\narrowtext

By interchanging the order of $x$ and $E$ integrations in Eqs. (6-7)
and performing the $x$-integration analytically, we see that in order
to keep the integrals finite, large powers of $1/E$ are not
allowed in the integrand. This requires that while using
 $\Phi_1$ the exponent should be bounded,
$\delta <3$, and while using $\Phi_2$ only the first three terms can be
present. Thus the model effectively has only two free parameters.

To determine $\mu_u$, $\mu_d$ and $T$, we solved the three coupled
nonlinear equations (6-7) by the Davidenko-Broyden method
\cite{antia}. The resulting values of $\mu_u$, $\mu_d$ and $T$ are
such that the left and right hand sides of these equations agree with
each other to typically one part in $10^6$.
The parton densities were evolved by means of the
Gribov-Lipatov-Altarelli-Parisi equations \cite{glap} in leading order
(LO), taking the input scale $Q^2_0=M^2$ and $\Lambda_{QCD}=0.3$ GeV.
Finally, the root-mean-square (rms) radius of the parton
distribution was taken to be the same as the charge rms radius
($\rho$) of the proton; since $\rho \simeq 0.862$ fm \cite{simon},
this yields $R = \sqrt{5/3}~ \rho = 1.11$ fm.

\medskip
\noindent {\it Results and Discussion}

Since the two arbitrary constants $B$ and $\delta$, or $a$ and $b$ in
Eq. (4) or (5) are not known, we have determined them by fitting the
deep inelastic scattering data on $F^p_2(x)$ at $Q^2=15$ GeV$^2$
\cite{hera,nmcbcdms}. The results of our fit incorporating the finite-size
corrections and QCD evolution are shown by the solid curve in
Fig. 2. (Results presented here are based on $\Phi_2$; the alternative
form $\Phi_1$ gives an equally good fit.) Also shown for comparison in
Fig. 2 are: (a) the (dot-dashed) curve labeled `GAS' giving the
prediction of the parton gas model which has no free parameters by
virtue of the constraints, (b) the (dashed) curve labeled `QCD'
showing the effect of QCD evolution on the gas model, and (c) the
(dotted) curve labeled `FSC' showing a {\it fit} to the data when only
the finite-size corrections are introduced in the gas model.
If $\Phi_1$ is used in order to incorporate FSC, the fitted values of
the two parameters are
\[B = 0.269 ~~\rm{and}~~ \delta = 2.14,\]
and the corresponding temperature and chemical potentials are $T
= 63$ MeV, $\mu_u = 124$ MeV and $\mu_d = 64$ MeV.
If, on the other hand, $\Phi_2$ is used,
the fitted values of the two parameters are \cite{amuse}
\[a = -1.88 ~~\rm{and}~~ b = 2.24,\]
and the corresponding temperature and chemical potentials are $T
= 72$ MeV, $\mu_u = 162$ MeV and $\mu_d = 81$ MeV.

To comment on the relative importance of the inputs, we focus on the
curves in Fig. 2 at, say, $x=10^{-3}$: a fit with FSC gives a very small
value of $F_2^p\sim 0.02$, reflecting the restrictive nature of the
constraints. Leading-order QCD evolution does result in a value of
$F_2$ which is significantly large but not large enough, $F_2^p\sim
0.23$. However, when the effects due to both FSC and QCD are included
in the model, we obtain $F_2^p\sim 1.1$, which is consistent with the
data. The presence of inverse powers of $(ER)$ in $\Phi$ is thus
partially responsible for increase in $F_2^p$ at small $x$.

As a test of the model, we show in Fig. 3 the prediction (solid curve)
for the difference $[F^p_2(x) - F^n_2(x)]$. Also shown for comparison
is the result (dashed curve) based on the parametrization of Gl\"uck
{\it et al.} \cite{params}. The agreement with the NMC data is
reasonable.

\newpage

As for the Gottfried sum $S_G$, we have
\begin{eqnarray}
S_G &=& {1 \over 3} - {2 \over 3} \int^1_0 (\bar d - \bar u) dx\nonumber\\
&=&{1 \over 3} - {2 \over 3}~{V M^2 \over (2\pi)^2} \int^1_0 dx ~x
\int^{M/2}_{xM/2} dE~ \bigg\{ {6 \over \exp \big[\beta (E + \mu_d)\big]
+ 1} - {6 \over \exp \big[\beta (E + \mu_u)\big] + 1}\bigg\}\Phi(ER).
\nonumber \\ \eqnum{8}
\end{eqnarray}
The inequality $S_G < {1 \over 3}$ is thus a result of
having in the proton, more valence $u$ quarks than valence $d$ quarks,
$(n_u - n_{\bar u}) > (n_d - n_{\bar d})$, implying that $\mu_u >
\mu_d$ and hence the integral in Eq. (8) is positive. Our model
predicts at $Q^2 =4$ GeV$^2$, the value $S_G=0.22$ which is
consistent with the experimental value $S_G=0.235\pm 0.026$.

The rapidity dependence of the $W$ charge asymmetry in the reactions
$\bar pp \rightarrow W^{\pm }~+~ ...$ is now known to a very high
precision \cite{cdf}. It is a sensitive function of the quark flavor
ratio $d(x)/u(x)$ in the proton, in the range $0.007 < x < 0.24$ at
$Q^2 = M_W^2$. The ratio $\bar u(x)/ \bar d(x)$ at $<x>=0.18$
has been deduced to be about $0.51$ by
the NA51 collaboration \cite{NA51}. These and other predictions of the
model, on the ratio $(F^n_2(x)/ F^p_2(x))$, the quark and antiquark
distributions $q(x)$, $\bar q(x)$, $q_v(x) = q(x) - \bar q(x)$ for
various flavors, the gluon distribution $g(x)$, the longitudinal
momentum fraction carried by the charged partons, etc. will be given
elsewhere \cite{rsbkvl}.

Now we briefly describe the salient features of some of the recent
calculations of the nucleon structure functions, which use ideas from
statistical mechanics.
Mac and Ugaz [8a] calculated first-order QCD corrections to the
statistical distributions and obtained a crude but reasonable
agreement with $F_2^p(x)$ data for $x \gsim 0.2$. The momentum
constraint was not imposed and the fitted value of the proton radius
$(R)$ was 2.6 fm.
Cleymans {\it et al.} [8b] used the framework of the finite
temperature quantum field theory. They considered ${\cal O}(\alpha_s)$
corrections to the statistical distributions and obtained a good fit
to the $F_2^p(x)$ data for $x \geq 0.25$. They also calculated the
ratio $\sigma_L / \sigma_T$ in this region; it was a factor of 6 above
the experimental value.
Bourrely {\it et al.} [8c] considered polarized as well as
unpolarized structure functions and presented a statistical
parametrization (with eight parameters) of parton distributions in the
IMF. Their framework allowed chemical potential for quarks as well as for
gluons. The number constraints were not satisfied very accurately. QCD
effects were not considered. $x \bar q(x)$ vanished as $x \rightarrow
0$ and so it was not possible to reproduce the fast increase of the
antiquark distributions for $x<0.1$.
Bourrely and Soffer's [8d] approach was similar to that in [8c]. By
incorporating QCD evolution of parton distributions and allowing the
antiquark chemical potential to depend on $x$, they were able to
reproduce the HERA data on $F_2^p$.

In conclusion, it is noteworthy that the application of ideas of
statistical mechanics to the point constituents of the nucleon can
provide a simple description of all the observed features of the
(unpolarized) nucleon structure functions down to the lowest $x$
values so far explored. The model presented here has two free
parameters which arise from our treatment of the finite-size
corrections.

\acknowledgments

I am grateful to K.V.L. Sarma for numerous discussions during the
course of this work and for a critical reading of the manuscript. I
thank R.M. Godbole and S. Kumano for discussions and communications
regarding QCD evolution. I benefited from valuable comments by
A.K. Rajagopal, Virendra Singh and C.S. Warke.

\begin{figure}
\caption
{Log-log plot of the proton structure function data.
Experimental data are from
%Refs. \cite{hera} and \cite{nmcbcdms};
Refs. [4,5]; the error bars show statistical and
systematic errors combined in quadrature. The straight lines
are our fits described in Eq. (1), and are labeled by $Q^2$ = 15,~35,
and 120 GeV$^2$. Numbers have been scaled by the factors shown in
parentheses for convenience in plotting.}
\label{1}
\end{figure}

\begin{figure}
\caption
{Proton structure function $F^p_2(x)$ at $Q^2 = 15$ GeV$^2$. Data points
are as in Fig. 1.
Solid curve is our best fit to the data. Also shown for comparison
are: the (dot-dashed) curve labeled `GAS' giving the gas model
prediction, the (dashed) curve labeled `QCD' showing the QCD-evolved
gas model, and the (dotted) curve labeled `FSC' which is a fit to the
data when finite-size corrections are included in the gas model
(without QCD).}
\label{2}
\end{figure}

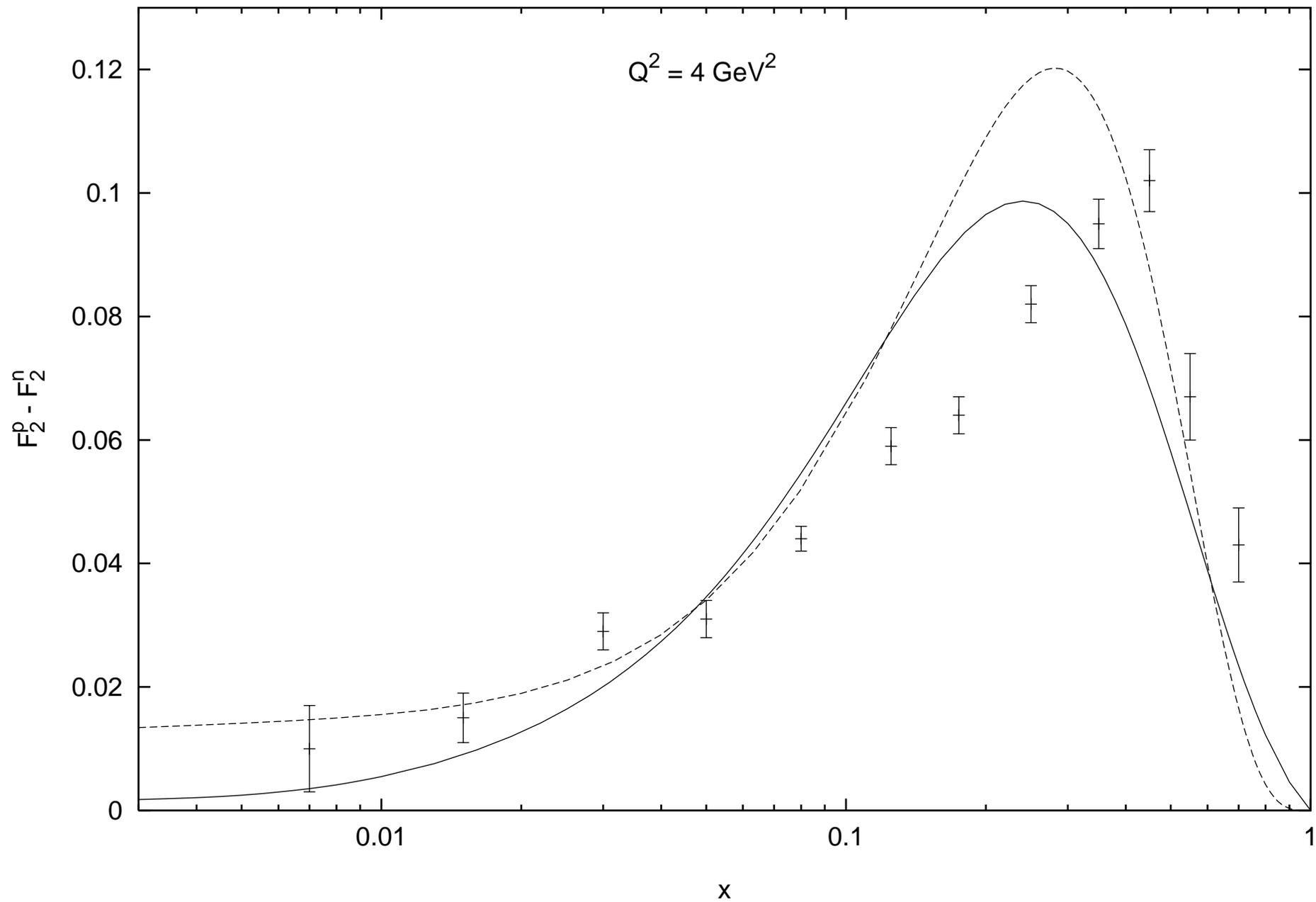
\begin{figure}
\caption
{Difference $(F^p_2-F^n_2)$ versus $x$, at $Q^2
= 4$ GeV$^2$. Experimental data are from
%Ref. \cite{nmcgsr};
Ref. [1];
errors are statistical only. Solid curve is the prediction of our model.
Dashed curve is based on the parametrization of Gl\"uck {\it et al.} [7].}
\label{3}
\end{figure}


\begin{references}
\bibitem{amuse} It is amusing to note that the values of $a(=
-1.88)$ and $b(= 2.24)$ determined by us are
close to the values $a=-3\pi/4$ and $b=9/4$
which follow from one of the expressions for $dn/dE$
given by Morse and Ingard \cite{fse}; substitute $ c = 1, \nu =
E/(2\pi), A=4 \pi R^2 $ and $\ell_x =\ell_y=\ell_z=2R$ in $dN_{ob}$ in
their Eq. (9.5.12).
\end{references}
\end{document}